\documentclass[aps,prl,twocolumn,showpacs,preprintnumbers]{revtex4}

\usepackage{psfrag,graphicx}
\usepackage{dcolumn}
\usepackage{amsmath,amssymb}
\usepackage{bm}
\usepackage{amsfonts,amssymb,amsmath}        

\newcommand{\be}{\begin{equation}}
\newcommand{\ee}{\end{equation}}
\newcommand{\bq}{\begin{eqnarray}}
\newcommand{\eq}{\end{eqnarray}}

\newcommand{\no}{\nonumber\\}

\newcommand{\ket}[1]{\left | \, #1 \right\rangle}
\newcommand{\bra}[1]{\left \langle #1 \, \right |}
\begin{document}

\title{Non-abelian statistics from an abelian model}

\author{James R. Wootton$^1$, Ville Lahtinen$^1$, Zhenghan Wang$^2$, Jiannis K. Pachos$^1$}
\affiliation{$^1$School of Physics and Astronomy, University of Leeds, Woodhouse Lane, Leeds LS2 9JT, UK\\
$^2$Microsoft Research, Station Q, University of California, Santa Barbara, CA 93106, USA}
\date{\today}

\begin{abstract}

It is well known that the abelian $Z_2$ anyonic model (toric code) can be realized on a highly entangled two-dimensional spin lattice, where the anyons are quasiparticles located at the endpoints of string-like concatenations of Pauli operators. Here we show that the same entangled states of the same lattice are capable of supporting the non-abelian Ising model, where the concatenated operators are elements of the Clifford group. The Ising anyons are shown to be essentially superpositions of the abelian toric code anyons, reproducing the required fusion, braiding and statistical properties. We propose a string framing and ancillary qubits to implement the non-trivial chirality of this model.

\end{abstract}


\maketitle

Anyons are two dimensional particles that, unlike boson or fermions, satisfy exotic statistics \cite{Wilczek,Witten,Pachos2,Mund,pachossup}. These are manifested for abelian anyons by a phase factor when they are interchanged or for non-abelian anyons by a unitary matrix \cite{Preskill}. This complex behavior makes it a challenging task to find a representation, mathematical or physical, that reproduces these properties. Recently, there has been increased interest in these models due to their connection to fault-tolerant quantum computation \cite{Kitaev2,Dennis}, and their relation to new states of topologically ordered matter \cite{Wen2}. They are also of interest in the study of multipartite entanglement, due to the so-called topological entanglement \cite{topent} required to realize anyons on physical systems. Several proposals have been made~\cite{Laughlin,Read,Freedman,Wen} for physical systems with anyonic behavior. Of particular interest are lattice models, where qubits are placed on a two dimensional surface with their states representing the vacuum or anyonic populations \cite{Kitaev,Kitaev2,Levin,Fendley,Brennen}. While for the quantum double models we have a simple spin representation of the corresponding states \cite{Kitaev2,Levin,Dijkgraaf,Bais}, it has been rather hard to identify the states of other models, such as the so-called Ising anyonic model. 

Recently, Kitaev~\cite{Kitaev} presented a spin lattice Hamiltonian that for different coupling regimes exhibits abelian or non-abelian anyonic behavior \cite{Lahtinen,Yu}. The former corresponds to the well studied toric code model, while the latter corresponds to the Ising anyonic model, whose properties have proven difficult to demonstrate. It has been shown in the context of the fractional quantum Hall effect that states of the Ising non-abelian anyons can be built up from states with abelian statistics \cite{Stern,Georgiev}. In this letter we demonstrate that the topologically entangled states of a spin lattice capable of supporting the anyons of the toric code are sufficient also to support the anyons of the Ising model. This is done by reproducing the states of the Ising model through quantum superpositions of the toric code states. Our aim is to present the mechanism that is responsible for their exotic properties without invoking the Hamiltonian description. In this scheme the Ising model anyons are located at the endpoints of strings of operators that belong in the Clifford group. Thus, for the first time we give a lattice representation of a non-abelian model that is not a quantum double. We achieve this by demonstrating that complex topological models can be constructed from simpler models, providing a general methodology to perform other such mappings.

\begin{center}
\begin{figure}[ht]
\resizebox{7cm}{!}
{\includegraphics[scale=1]{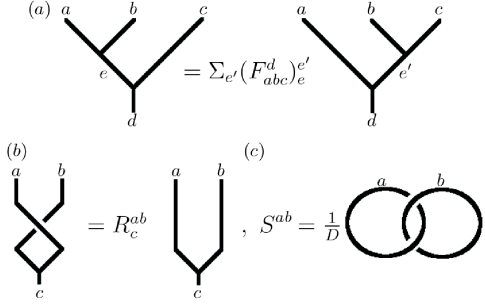} }
\caption{\label{fandr}The (a) $F$ matrices, (b) braidings $R$ and (c) $S$-matrix of a general anyonic model. The vertical axis represents time running downwards.}
\end{figure}
\end{center}

To define an anyonic model one needs to give a set of possible particle types and their fusion rules. For abelian models these rules take the form $a \times b = c$, with only one outcome possible for each fusion. For non-abelian models, however, the rules take the more general form $a \times b = \sum_c N_{ab}^c c$, with multiple possible fusion outcomes. These rules also determine the quantum dimension, $d_a$, of each particle, $a$, with the total quantum dimension of the model defined by $D^2=\sum_{a} d_a^2$. Further, one needs the $R$- and $F$-matrices that describe the action of a counter clockwise exchange of particles and changing the fusion order, respectively (see Fig. \ref{fandr}(a,b)). These must satisfy the so-called pentagon and hexagon relations which restrict to a finite set of consistent theories \cite{oceanu}. These concepts can be summarized using the topological $S$-matrix,
\be
S^{ab} = \frac{1}{D}\sum_c d_c \textrm{tr}(R^{ab}_c R^{ba}_c),
\label{s}
\nonumber
\ee
which can be interpreted as the vacuum-to-vacuum process depicted in Fig. \ref{fandr}(c).

\vspace{-.25cm}
\vspace{-.85cm}

{\em Toric code and Ising models:-} The toric code model consists of four different particle types, the vacuum, $1$, the anyons $e$ and $m$ and the fermion $\epsilon$. The non-trivial fusion rules for these particles are 
\be
e \times e = m \times m = \epsilon \times \epsilon = 1, \,\,
e \times m = \epsilon, \,\,
e \times \epsilon = m, \,\,
m \times \epsilon = e.
\nonumber
\ee
The quantum dimensions are $d_a=1$ for all $a$, hence $D^2=4$. The $R$ matrices of interest to us are given by
\be
R^{\epsilon \epsilon}_{1} = (R^{e m}_{\epsilon})^2 = -1, \,\,
 R^{e e}_{1} = R^{m m}_{1} = 1,
\nonumber
\ee
and all of the $F$-matrices are equal to the identity.

It is possible to represent these anyons on a honeycomb lattice with qubits placed on the vertices. Each plaquette $P$ is split into two subplaquettes, labeled $s$ to the left and $p$ to the right, in order to facilitate the implementation of the two different species of anyons. We identify with the vacuum a state satisfying $A_s \ket\xi = \ket\xi$ and $B_p \ket\xi = \ket\xi$ for all $s$ and $p$, where $A_s=\sigma^x_1\sigma^x_2\sigma^x_3\sigma^x_4$ and $B_p=\sigma^z_1\sigma^z_2\sigma^z_3\sigma^z_4$ are products of Pauli matrices acting on the four qubits of each subplaquette. States having any other pattern of eigenvalues can then be identified with anyonic populations. For example, an $e$ anyon pair is given by the state, $\ket{e, e}=\sigma^z_i\ket{\xi}$, corresponding to a $-1$ eigenvalue of the $A_s$ on the $s$ subplaquettes neighboring site $i$. A string of $\sigma^z$'s corresponds to two $e$ anyons positioned at its endpoints. Similarly the endpoints of $\sigma^x$ and $i\sigma^y$ strings are $m$ and $\epsilon$ anyons, respectively. All the properties of these anyons can easily be reproduced from the representation of their corresponding strings by the Pauli operators \cite{Kitaev3}.

On the other hand, the Ising model consists of three different particle types, the vacuum, $1$, the non-abelian anyon $\sigma$ and a fermion, $\psi$. The non-trivial fusion rules are
\be
\sigma \times \sigma = 1 + \psi, \,\,
\psi \times \psi = 1, \,\,
\sigma \times \psi = \sigma.
\nonumber
\ee
Again all particles are their own antiparticles. The quantum dimensions are given by $d_1=d_{\psi}=1$ and $d_{\sigma}=\sqrt{2}$, which also leads to $D^2 = 4$. The $R$ matrices of interest to us are
\bq \nonumber
R_1^{\psi \psi} = -1&,& \,\,
(R_{\sigma}^{\psi \sigma})^2 = -1, \\
(R_1^{\sigma \sigma})^2 = e^{-i\pi/4}&,& \,\,
(R_{\psi}^{\sigma \sigma})^2 = - e^{-i\pi/4},
\nonumber
\eq
This model has the non-trivial $F$ matrix 
\be
F_{\sigma \sigma \sigma}^{\sigma} = \frac{1}{\sqrt{2}}\left(
\begin{array}{rr}
1 & 1 \\
1 & -1 
\end{array} \right), \label{f}
\ee
in the basis $1$, $\psi$. This shows that a change in the fusion order of $\sigma$ particles leads to a superposition of different fusion outcomes.

{\em Superposition Principle:-} There is a simple argument that shows a relation between the toric
code and the Ising models. Consider their corresponding $S$-matrices, given by
\bq
S_{Z_2} &=& \frac{1}{2}\left(
\begin{array}{rrrr}
1 & 1 & 1 & 1 \\
1 & 1 & -1 & -1 \\
1 & -1 & 1 & -1\\
1 & -1 & -1 & 1\\
\end{array}
\right)\!, \nonumber \\
S_\text{Ising} &=& \frac{1}{2}\left(
\begin{array}{ccc}
1 & \sqrt{2} & 1 \\
\sqrt{2} & 0 & -\sqrt{2}\\
1 & -\sqrt{2} & 1\\
\end{array}
\right),
\nonumber
\eq
defined in the basis, $1$, $e$, $m$, $\epsilon$ and $1$, $\sigma$, $\psi$,
respectively. Let us consider the equal superpositions of the $e$ and $m$ particle
loops of the toric code. Then the following relations hold
\bq
S_\text{Ising}^{11}&=& S_{Z_2}^{11}=1, \,\,
S_\text{Ising}^{\psi\psi}=S_{Z_2}^{\epsilon\epsilon}=1,
\no
S_\text{Ising}^{1\sigma}&=&{S_{Z_2}^{1e}+S_{Z_2}^{1m} \over
\sqrt{2}}=\sqrt{2},
\no
S_\text{Ising}^{\psi\sigma}&=&{S_{Z_2}^{\epsilon e}+S_{Z_2}^{\epsilon m}
\over\sqrt{2}}=-\sqrt{2},
\no
S_\text{Ising}^{\sigma\sigma}&=&{S_{Z_2}^{ee}+S_{Z_2}^{em}
+S_{Z_2}^{me}+S_{Z_2}^{mm}\over 2} =0.
\eq
Further, note that $d_{\sigma}^2 = d_e^2+d_m^2$ and that the total quantum dimensions of the two models are equal \cite{qd}.

These observations motivate us to identify a $\sigma$ particle loop of the Ising anyon model with these superposed $e$ and $m$ loops, and to identify the fermions of the two models. Using this superposition principle we can demonstrate fusion, braiding and statistical characteristics of the Ising anyon model. In addition, we employ an ancillary system that provides the correct chirality, and will at times require its state to be an entangled state with the lattice.

\begin{center}
\begin{figure}[ht]
\resizebox{8cm}{!}
{\includegraphics{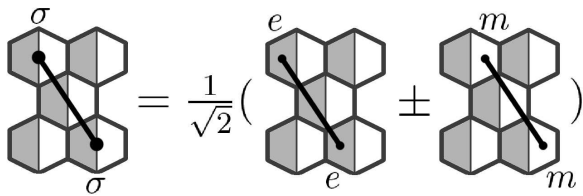} }
\caption{\label{superposition}The state of a $\sigma$ string with endpoints in two plaquettes of the honeycomb lattice can be described by a superposition of $e$ and $m$ strings. The relative $\pm$ sign is a non-local property that cannot be accessed by measurements at either endpoint.}
\end{figure}
\end{center}

The state of a superposition of an $e$ string and an $m$ string which end in the same plaquettes, may be written as
\be
\ket{\sigma_1, \sigma_2 ; j} = \frac{1}{\sqrt{2}}(\ket{e_1, e_2} + j \ket{m_1, m_2}),
\ee
where, for example, we have used $\ket{e_1, e_2}$ to denote the state of an $e$ string with one endpoint a plaquette labeled $P=1$ and the other in $P=2$ (Fig. \ref{superposition}). This can be viewed as a new string whose endpoints can reproduce the behavior of the $\sigma$ anyons, and so have been labelled as such. The relative sign $j$ is a non-local property of the string which cannot be determined by local observations of the endpoints. It can be changed by braiding operations which may act locally around endpoints, and so is not topologically protected. The fermions of the toric code will reproduce the behaviour of those of the Ising model, so we will identify $\epsilon$ strings with $\psi$ strings.

The movement of a $\sigma$ must be performed in such a way that it does not affect the superposition that encodes the non-abelian character of the anyons \cite{Aguado}. This can be done by using a qubit ancilla, initially in state $\ket{0}_q$, and the controlled operations
\bq
C_s &=& \frac{1}{2}(\openone + A_s)\otimes \openone_q + \frac{1}{2}(\openone - A_s)\otimes \sigma^x_q , \no
D_i &=& \sigma_i^x \otimes \ket{0}\bra{0}_q + \sigma_i^z \otimes \ket{1}\bra{1}_q .
\eq
Applying $C_s$ entangles the $s$ plaquette at the endpoint of a $\sigma$ string with the ancilla. The operation $D_i$ may then be applied between the ancilla and the lattice qubit $i$ to extend the string one step. To unentangle the ancilla $C_{s'}$ is applied, using the plaquette $s'$ at the new endpoint of the $\sigma$ string. This method of extending the strings is local, allowing the interpretation of their endpoints as particles. Also, these operators reproduce the braiding statistics of the toric code.

\vspace{-.25cm}
\vspace{-.85cm}

Because we represent particles as endpoints of strings, we do not consider any process which cannot be described purely in terms of them. So we restrict ourselves to only considering the fusion processes in the Ising model that can be thought of as two strings fusing to form another. This process will be referred to as the fusion of strings. We further restrict that the composite string belongs to the vacuum sector. As an example, the state of two $\sigma$ strings can be written as
\bq
&&\ket{(\sigma_1, \sigma_2 ; j)(\sigma_3, \sigma_4 ; k)} =
\no&&\,\,\,\,\,\,\,\, \frac{1}{2}(\ket{e_1, e_2, e_3, e_4} + jk\ket{m_1, m_2, m_3, m_4} +
\no &&\,\,\,\,\,\,\,\,\,\,\,\,\,\,\,\,\, k \ket{e_1, e_2, m_3, m_4} + j \ket{m_1, m_2, e_3, e_4}),
\label{supstate}
\eq
where the relative signs are given by $j,k \in \{-1,+1\}$. The fusion of the strings is achieved by the fusion of the particles residing at plaquettes $1$ and $3$ and of those at $2$ and $4$ \cite{note}. The endpoints in the first two terms will each behave as the vacuum since each composite object is made up of either two $e$'s or two $m$'s, and similarly the second two terms will each give a fermion string. If $j=k$ the result of the fusion is then
\be
\ket{(\sigma_1, \sigma_2 ; j)(\sigma_3, \sigma_4 ; j)} =\frac{1}{\sqrt{2}}(\ket{1_{1,3}, 1_{2,4}} + j \ket{\psi_{1,3} \psi_{2,4}}),
\ee
where we have used $\ket{1_{1,3}, 1_{2,4}} = (\ket{e_1, e_2, e_3, e_4} + \ket{m_1, m_2, m_3, m_4})/\sqrt{2}$ to denote the terms that fuse to vacuum and $\ket{\psi_{1,3} \psi_{2,4}} = (\ket{e_1, e_2, m_3, m_4} + \ket{m_1, m_2, e_3, e_4})/\sqrt{2}$ to denote the terms that fuse to a fermion string.

Let us identify $\ket{\sigma, \sigma ; \pm}$ with pairs belonging to the vacuum and fermion sectors, respectively. The above result then reproduces the $F$ matrix (\ref{f}), and therefore the fusion properties, of the Ising anyon model. The fusion of a $\sigma$ string belonging to the vacuum sector with one belonging to the fermion sector does not result in a composite string, and so we need not consider cases where $j \neq k$.

By considering the decompositions of the $\sigma$ and $\psi$ particles in terms of the toric code particles we can show that they satisfy the Ising model braiding rules. For example, let us consider the exchange of two $\psi$'s. Since these are identified with the $\epsilon$'s of the toric code they will have the same fermionic behavior. Also, since the braiding of an $e$ or an $m$ around an $\epsilon$ results in a phase factor of $-1$, so does the braiding of a $\sigma$ around a $\psi$. Let us also consider the braiding of two of the $\sigma$ particles, such as those in Eq. (\ref{supstate}). Braiding the $\sigma$ residing at plaquette $1$ around that at $3$  results in a change of the relative sign for both $\sigma$ strings, and so a change also of the relative sign between the vacuum and fermion strings in the fusion outcome. From this we infer the $R$ matrices $(R_1^{\sigma \sigma})^2 = 1$ and $(R_{\psi}^{\sigma \sigma})^2 = - 1$. These are similar to those of the Ising model, except that a complex phase factor is missing. This required phase differs for counter clockwise and clockwise braidings, $e^{-i\pi/4}$ for the former and $e^{i\pi/4}$ for the latter. Since $R=R^{\dagger}$ for the toric code particles, the lattice does not distinguish between counter clockwise and clockwise evolutions. A framing \cite{Levin} is therefore proposed for the $\sigma$ particles to make this distinction and to encode the chirality on an ancillary system.

\begin{center}
\begin{figure}[ht]
\resizebox{4.5cm}{!}
{\includegraphics{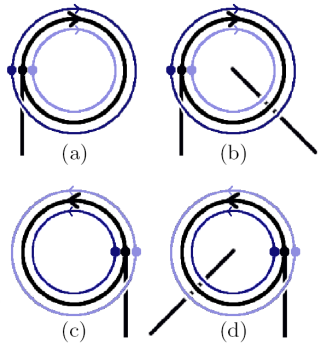}}
\caption{\label{loops}(Color online) The four possible loops for a $\sigma$ particle that start and finish at the marked points, where the framing is depicted. The loops to the top are clockwise and those to the bottom are counter clockwise . The loops to the left enclose no other $\sigma$ particle and those to the right do. In (a) and (d) the left framing crosses a $\sigma$ string once and the right framing does not cross or crosses twice, resulting in a bit flip on the ancillary qubit for the left framing only. In (b) and (c) the situation is reversed.}
\end{figure}
\end{center}

We allocate two framings to each $\sigma$ particle, one to the left ($l$) and one to the right ($r$). Each of them has an ancillary qubit, initially in the zero state, $\ket{0}_l \ket{0}_r$. When the particle moves the framings move with it, performing the operation
\be
E_i = \openone_i \otimes \ket{+}\bra{+} + i \sigma_i^y \otimes \ket{-}\bra{-}.
\ee
between their ancillary qubits and the lattice sites, $i$, to the left and right of the particle. This creates superpositions of the vacuum and a fermion on the lattice, controlled on the ancilla state. When the loops are complete the framings act trivially on the lattice, but may cause a bit flip on the ancilla depending on whether the fermion loop acquired a $-1$ by crossing a $\sigma$ string. After each loop the ancillary qubits are measured and the operations $e^{i \pi/8} \sigma^x_r$ and $e^{-i \pi/8} \sigma^x_l$ applied for the results $\ket{0}_l \ket{1}_r$ and $\ket{1}_l \ket{0}_r$ respectively. These assign a phase and reset the qubits. The state $\ket{0}_l \ket{1}_r$, for example, is assigned $e^{i \pi/8}$ since it is the result of either a counter clockwise loop that encloses no other $\sigma$ particle or a clockwise loop which does enclose a $\sigma$ particle. In the former case this phase comes from the fact that the loop causes the extended object of the $\sigma$ particle and framing to undergo a counterclockwise twist of $2 \pi$. This must therefore be assigned the phase $e^{i \pi/8}$, due to a topological spin. In the latter case the phase comes from both a clockwise braiding and a twist, $e^{i \pi/4}e^{-i \pi/8}=e^{i \pi/8}$. The consistency of this framing can be verified in Fig. \ref{loops}, where a complete set of elementary cases have been considered.

The phase factor required for the $R$ matrix is that for a braiding in which a $\sigma$ particle performs a loop around another particle without twisting. So the twists must be removed from the above loops in order to obtain the corresponding evolutions. This can be done by following all loops with a twist alone in the opposite direction. By this two stage process the framing applies the required phase of $e^{-i\pi/4}$ for a counter clockwise braiding. When the phase is inserted it gives the $R$ matrix required for the consistency of the Ising model.

\vspace{-.25cm}
\vspace{-.85cm}

We may define the plaquette operator $W_P = A_s B_p$ on the plaquettes of the honeycomb lattice, where $s$ and $p$ are the subplaquettes of $P$. This detects whether an $e$ or an $m$ is present on $P$ without distinguishing between the two. It can therefore detect the presence of a $\sigma$ particle as defined above, without collapsing or otherwise changing the superposition.  The $W_P$'s defined in this way are equivalent to those of Kitaev's honeycomb lattice model \cite{Kitaev}. This means that $e$ and $m$ particles and their superpositions correspond to the vortex-like excitations of this model, as one would expect for the Ising $\sigma$'s. Similarly the fermions, which are not detected by $W_P$, correspond to non-vortex like excitations.

{\em Conclusions:-} In this letter we demonstrated that the superposition of the states of the toric code, together with the appropriate framing, can reproduce the fusion, braiding and statistical properties of the Ising model. This is a surprising connection between an abelian and a non-abelian anyonic model that reveals the non-local character of the latter. It also gives the first lattice  representation of a non-abelian model that is not a quantum double. It is an exciting possibility to verify if such a relation holds between other models, to derive a Hamiltonian that has these states at its low energy spectrum or to implement these non-abelian states in the laboratory \cite{Pachos,Yang}.

{\em Acknowledgements:-} We would like to thank Joost Slingerland and Xiao-Gang Wen for inspiring conversations. This work was supported by the EU grants SCALA and EMALI, the EPSRC, the Finnish Academy of Science and the Royal Society.


\begin{thebibliography}{99}
\bibitem{qd} This is an important point, since the topological entanglement entropy of states that support a given anyon model depends on the total quantum dimension. So if we want to realise two models with the same topologically entangled states, it seems natural to take the total quantum dimensions to be equal.
\bibitem{note} This does not necessarily mean that we need to move these particles to the same plaquettes, only that we have to consider the braiding properties of the composite objects.
\end{thebibliography}
\end{document}